\documentclass[osajnl,twocolumn,showpacs,superscriptaddress,10pt]{revtex4-1}
\usepackage{amsmath,amssymb,graphicx}
\begin{document}

\title{General Approach for the Sensitivity Analysis and Optimization of Integrated Optical Evanescent-Wave Sensors}

\author{Camille Delezoide}\email{Corresponding author: camille.delezoide@lpqm.ens-cachan.fr}
\author{Isabelle Ledoux-Rak}
\author{Chi Thanh Nguyen}
\affiliation{LPQM, Institut d'Alembert, Ecole Normale Sup\'erieure de Cachan, \\ 61, avenue du Pr\'esident Wilson, 94230 Cachan, France}


\thispagestyle{empty}
\setcounter{page}{0}

This paper, “General approach for the sensitivity analysis and optimization of integrated optical evanescent-wave sensors”, was published in Journal of the Optical Society of America (JOSA) B and is made available as an electronic reprint with the permission of OSA. The paper can be found at the following URL on the OSA website: http://dx.doi.org/10.1364/JOSAB.31.000851. Systematic or multiple reproduction or distribution to multiple locations via electronic or other means is prohibited and is subject to penalties under law.

\clearpage
\begin{abstract}
The optimization of integrated optical evanescent-wave sensors is dual. For optimal performances, we require waveguides with both maximal sensitivity to the measurand, the quantity intended to be measured, and minimal sensitivities to perturbations. In this context, fully numerical approaches are extremely powerful, but demand huge computer resources. We address this issue by introducing a general and efficient approach, based on the formal derivation of analytical dispersion equations, to express and evaluate all waveguide sensitivities. In particular, we apply this approach to rectangular waveguides, to discuss its accuracy and its use within sensitivity optimization procedures.
\end{abstract}


\maketitle 

\section{Introduction}\label{sec:Intro}

A wide variety of photonic devices integrate optical waveguides \cite{Lifante03,Ma02}, and waveguide design often plays a key role in the performances of these devices. For optimization purposes, numerous studies were led to model, both analytically \cite{Marcatili69}, \cite{Hocker77}, \cite{Kumar83} and numerically \cite{Goell69}, \cite{Scarmozzino00}, the guided optical modes inside these waveguides \cite{Okamoto00}. For typical waveguide structures, such as rectangular waveguides, the use of common analytical models results in a limited accuracy on the optimization functions to be evaluated, especially for waveguides with small cross-sections. For this reason, analytical models have been progressively replaced by powerful numerical simulation tools.

Meanwhile, various waveguide structures were proposed as transducers in the domain of evanescent-wave biochemical sensors \cite{Fan08}, with very encouraging results. In these sensors, variations of the measurand - the quantity intended to be measured - are directly converted into variations of effective indexes \cite{Okamoto00} for guided modes propagating in the sensitive part of the device. It can be, for instance, the sensing arm of a Mach-Zehnder interferometer (MZI) \cite{Heideman93} or an entire optical microresonator \cite{Vollmer08, Delezoide12}. As a consequence, for such applications, a central issue is to optimize the sensitivity of modal effective indexes to the measurand. Another important issue is to minimize the sensitivities of modal effective indexes to perturbative quantities; typically, the temperature of the integrated device. This is actually a matter of concern for numerous waveguide applications besides biochemical sensors.

Both issues can be addressed using analytical or numerical waveguide models to calculate modal effective indexes for a large number of parameter sets. An optimal waveguide configuration is thus found using discrete derivatives to deduce waveguide sensitivities from calculated effective index values. However, this straightforward approach to waveguide sensitivity analysis is not optimal as far as computing time is concerned.

In this article, we introduce, test and discuss a more efficient approach to waveguide sensitivity analysis. It is based on the formal derivation of dispersion equations - provided by analytical models - on a neighborhood of a given set of parameters describing the waveguide. In Part 2, we illustrate how our approach works by applying it to the simple case of the slab waveguide. In most of the article though, we focus rectangular waveguide sensitivity analysis. First, in Part 3, we apply our approach to express the sensitivities of rectangular waveguides using two simple analytical methods, the Effective Index Method (EIM) and Marcatili's Method (MM). Thanks to these expressions, we calculate in Part 4 sensitivity values for the fundamental modes of a polymeric rectangular waveguide. These results are compared with reference values obtained from the Alternating Direction Implicit (ADI) numerical method using OptiMode, a state-of-the-art commercial mode solver. We then discuss the accuracy of our approach when combined with EIM and MM. In Parts 5 and 6, we apply our approach to perform quick trend analysis of rectangular waveguide sensitivities. More specifically, in Part 5, we apply our approach to study the sensitivity to cladding refractive index as quantity to maximize in evanescent-wave sensors. In Part 6, we study the waveguide thermal sensitivity, as quantity to minimize in most applications. In the rest of the article, we refocus on theoretical considerations related to our approach. In Part 7, by analyzing the computing time related to waveguide sensitivity analysis, we highlight the efficiency of our approach in comparison to what is achievable with a fully numerical approach.
Finally, in Part 8, we discuss the generalization of our approach to all waveguides.
\section{Effective index sensitivities of the slab waveguide}\label{sec:ImplSlab}
\subsection{Sensitivity coefficients}
\begin{figure}[ht]\begin{center}
\includegraphics [width=6cm]{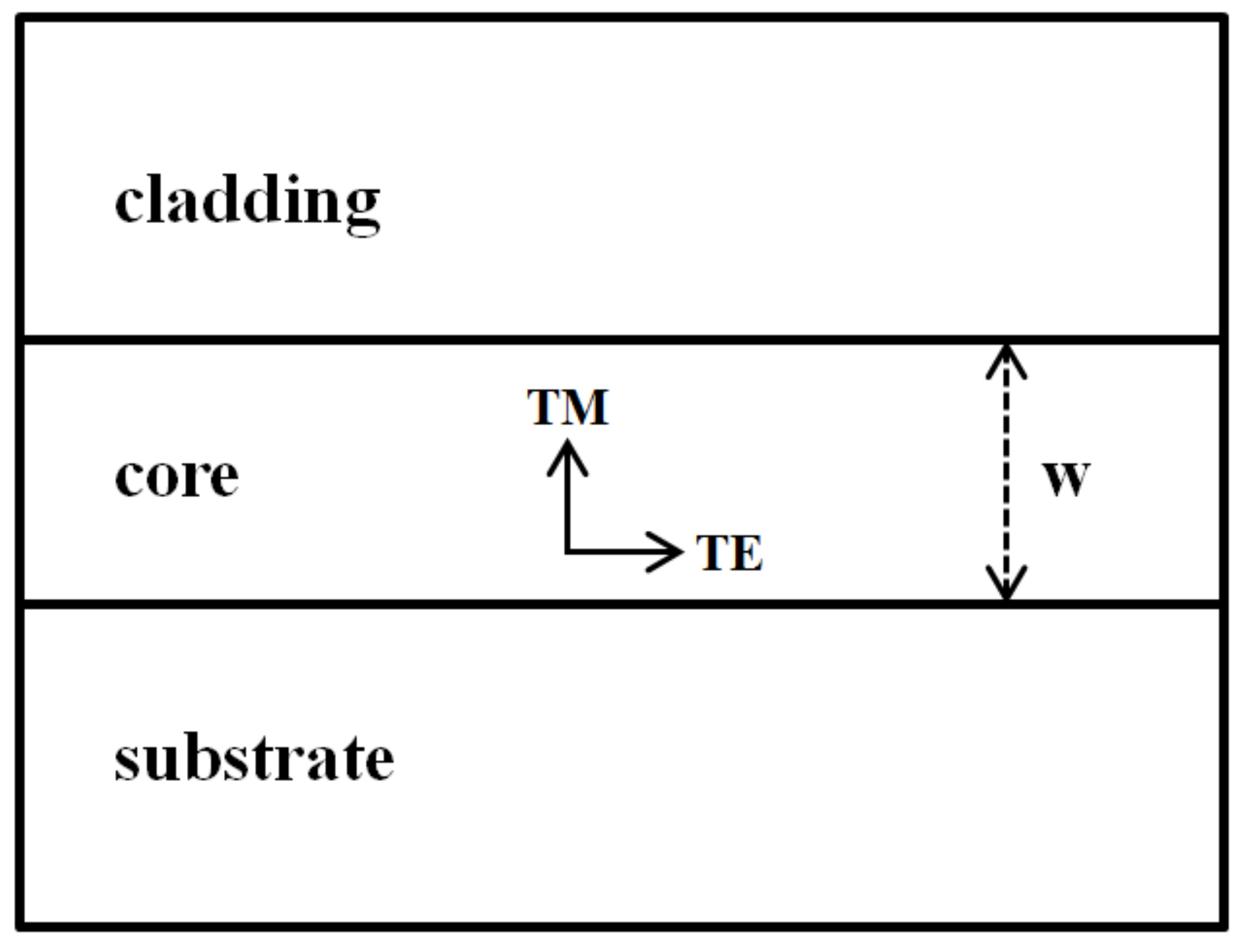}
\caption{Schematic drawing of an asymmetrical slab waveguide. Directions of the electric fields corresponding to TE and TM modes are represented.}
\label{fig:SlabWG}
\end{center}\end{figure}

As explained in Annex A, all dispersion equations for the slab waveguide (cf. Fig.\ref{fig:SlabWG}) can be written as identically zero functions $f(N_{eff},N_{core},N_{sub},N_{clad},w,\lambda)=0$ where $N_{eff}$, $N_{core}$, $N_{sub}$ and $N_{clad}$ are the effective, the core, the substrate and the cladding indices, respectively. $w$ is the width of the core  and $\lambda$ stands for the wavelength in vacuum.
Each function $f$ relates to a single guided mode with a given polarization and a given mode-order. 

First, we derive $f$ to obtain $df$, also identically zero:
\begin{equation}\label{3f}
\begin{split}
 df &= 0 =\frac{\partial f}{\partial \lambda}d\lambda+\frac{\partial f}{\partial N_{eff}} dN_{eff}+\frac{\partial f}{\partial N_{core}} dN_{core} \\
   &+\frac{\partial f}{\partial N_{sub}} dN_{sub} + \frac{\partial f}{\partial N_{clad}} dN_{clad}+\frac{\partial f}{\partial w}dw\\
\end{split}
\end{equation}
After simple arithmetical operations, we express the total derivative $dN_{eff}$ as a function of the sensitivity coefficients $S_i$ and of the total derivatives of waveguide parameters and wavelength:
\begin{equation}\label{3g}
\begin{split}
dN_{eff} &= S_{core}dN_{core}+S_{sub}dN_{sub} \\
   & \quad \quad+S_{clad}dN_{clad}+S_{w}dw+S_{\lambda}d\lambda
\end{split}
\end{equation}
 The sensitivity coefficients $S_i$ in Eq.\eqref{3g} are expressed as:
\begin{equation}\label{3h}
\left\{
\begin{aligned}
&S_{core} =\frac{\partial N_{eff}}{\partial N_{core}} =-\left(\frac{\partial f}{\partial N_{eff}} \right)^{-1}\frac{\partial f}{\partial N_{core}} \\
&S_{sub} =\frac{\partial N_{eff}}{\partial N_{sub}} =-\left(\frac{\partial f}{\partial N_{eff}} \right)^{-1}\frac{\partial f}{\partial N_{sub}}  \\
&S_{clad} =\frac{\partial N_{eff}}{\partial N_{clad}}= -\left(\frac{\partial f}{\partial N_{eff}} \right)^{-1}\frac{\partial f}{\partial N_{clad}} \\
&S_{w} =\frac{\partial N_{eff}}{\partial w}= -\left(\frac{\partial f}{\partial N_{eff}} \right)^{-1}\frac{\partial f}{\partial w} \\
&S_{\lambda} =\frac{\partial N_{eff}}{\partial \lambda}= -\left(\frac{\partial f}{\partial N_{eff}} \right)^{-1}\frac{\partial f}{\partial \lambda}
\end{aligned}
\right.
\end{equation}
Using the expressions above, evaluating all sensitivity coefficients $S_i$ for a given waveguide configuration ${\bf V_0} = (N_{core0},N_{clad0},N_{sub0},w_0,\lambda_0)$ comes down to:
\begin{enumerate}
\item Expressing the gradient $\nabla f$ as a function of $N_{eff}$, $N_{core}$, $N_{sub}$, $N_{clad}$, $w$, and $\lambda$.
\item Solving $f(N_{eff0},{\bf V_0})=0$ to obtain the value of $N_{eff0}$ related to the configuration ${\bf V_0}$.
\item Using $N_{eff0}$ and ${\bf V_0}$ to calculate the values of all components of $\nabla f$, then to deduce the values of sensitivity coefficients $S_i$ using Eq.\eqref{3h}. 
\end{enumerate}

\subsection{Sensitivity to a quantity X}
Thanks to the sensitivity coefficients $S_i$, we easily express the sensitivity $S_X$ of the modal effective index $N_{eff}$ to any quantity $X$:
\begin{equation}\label{3m}
\begin{split}
S_X &= \frac{dN_{eff}}{dX}= S_{core}\cdot\frac{dN_{core}}{dX}+S_{sub}\cdot\frac{dN_{sub}}{dX}\\
   & \quad+S_{clad}\cdot\frac{dN_{clad}}{dX}+S_{w}\cdot\frac{dw}{dX}
\end{split}
\end{equation}
$S_\lambda$ does not intervene here because $\lambda$ is an immutable parameter of the propagating field.

In order to evaluate the sensitivity $S_X$, the first task is to retrieve adequate values for coefficients ($d\cdot/dX)$, either from independent measurements or from the literature. Then, by repeatedly applying the three-step procedure described above to calculate sensitivity coefficients $S_i$ for various configurations, related values for the global sensitivity $S_X$ are easily deduced from Eq.\eqref{3m}. This method is applied in Part 6 to study the thermal sensitivity $S_T$ of a polymeric rectangular waveguide.

\subsection{Remarks}

We point out that in order to evaluate the sensitivity coefficients of a new configuration ${\bf V_0}$, only Steps 2 and 3 need to be repeated. This aspect is important for an optimal consumption of computer resources, as discussed in Part 7.

When implementing the three-steps procedure described in this part, a formal calculation engine is very handy to process the lengthy expressions of $\nabla f$ components. Using this method, the expression of $S_{clad}$ we obtain for a slab waveguide is identical to that published by Parriaux {\it et al.} \cite{Parriaux98}. Our approach can therefore be viewed as a generalization of their work, since it is virtually applicable to all waveguide sensitivities, and to all waveguides (cf. Part 8).

\section{Case of the rectangular waveguide}\label{sec:ImpRectWav}

\begin{figure}[ht]\begin{center}
\includegraphics [width=6cm]{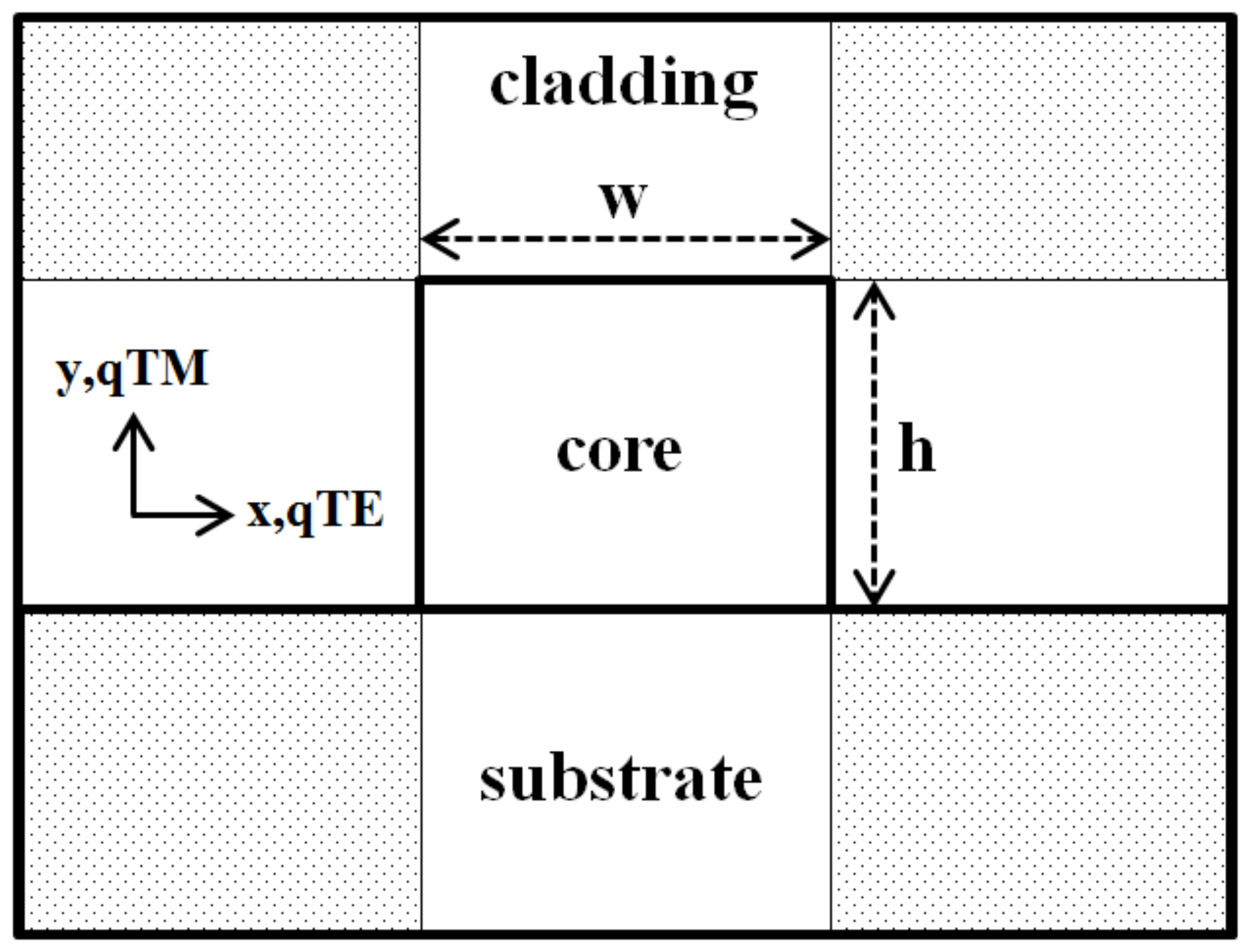}
\caption{Schematic drawing of a rectangular waveguide. Directions of the major electric field components corresponding to quasi-TE and quasi-TM modes are represented. Corner regions appear as shaded zones.}
\label{fig:RectWG}
\end{center}\end{figure}

The analytical method used to obtain dispersion equations for the slab waveguide, written as identically-zero $f$ functions in Part 2, is always the same as that described in \cite{Okamoto00}. For rectangular waveguides (Fig.\ref{fig:RectWG}) however, several analytical methods are typically used, and each method provides its own set of dispersion equations. Since our approach relies on dispersion equations to express the sensitivity coefficients $S_i$, all expressions, and accuracy of the values they generate, depend on the chosen analytical method (cf. Part 4).

\subsection{Analytical methods applicable to rectangular waveguides}
 
For rectangular waveguides, it is necessary to make assumptions in order to obtain analytical dispersion equations. These assumptions vary in form and number from one analytical method to another. As a result, analytical methods vary in accuracy \cite{Okamoto00}, which is important to consider when choosing a method. Also, some analytical methods are more general than others.



Amongst existing methods, Marcatili's Method (MM) is simple but only applicable to rectangular waveguides, whereas the Effective Index Method (EIM) is applicable to several other waveguides. A third method developped by Kumar \cite{Kumar83} uses a perturbation approach to deal with corner regions of the waveguides, resulting in a better accuracy than MM and EIM. However, the complex formalism developed in Kumar's method makes it laborious to use in our approach. As a consequence, we focus in this article on the use of EIM and MM to study rectangular waveguide sensitivities.

\subsection{Dispersion equations and sensitivity coefficients}

In Annexes B and C, we show that in both EIM and MM, each guided mode of a rectangular waveguide verifies a mode-order and polarization-dependent set of two dispersion equations written as identically zero functions $f_L(p)$ and $f_V(q)$. Similarly to the case of the slab waveguide, we also show how expressions of the sensitivity coefficients $S_i$ are deduced from the derivation of $f_L(p)$ and $f_V(q)$.

\section{Comparison with results from a numerical method}
\subsection{Methodology}
We choose here to study the fundamental TE and TM modes of a 2 $\mu m$-large polymer waveguide, with a varying height $h$. The core material is SU-8 \cite{Feng03,Tung05}, the substrate material is silica \cite{Malitson65} and the cladding material is water \cite{Kamikawachi08}. The wavelength is set at 1550 nm. 

We first compare the effective indexes obtained using both analytical methods, EIM and MM (cf. Annexes B and C), to the effective indexes obtained with the ADI numerical method using a state-of-the-art mode commercial mode solver, OptiMode.

Then, we proceed in the same fashion for $S_{clad}$, the sensitivity of the effective index to the cladding index $N_{clad}$. The values related to EIM and MM are obtained using a three-step procedure - identical to that described in Part 2 - to evaluate the expressions of $S_{clad}$ found in Annexes B and C. As for the ADI values, they are obtained by comparing the values of the effective indices of the TE and TM modes when $N_{clad}$ undergoes a slight variation - $\delta=0.001$ - of its value. Each value of the sensitivity $S_{clad}$ is then deduced from:
\begin{equation}\label{4Ca}
S_{clad}^{ADI}=\frac{N_{eff}^{ADI}\left( N_{clad}+\delta \right)-N_{eff}^{ADI}\left( N_{clad} \right)}{\delta}
\end{equation}

\begin{figure}[here]\begin{center}
\includegraphics [width=8cm]{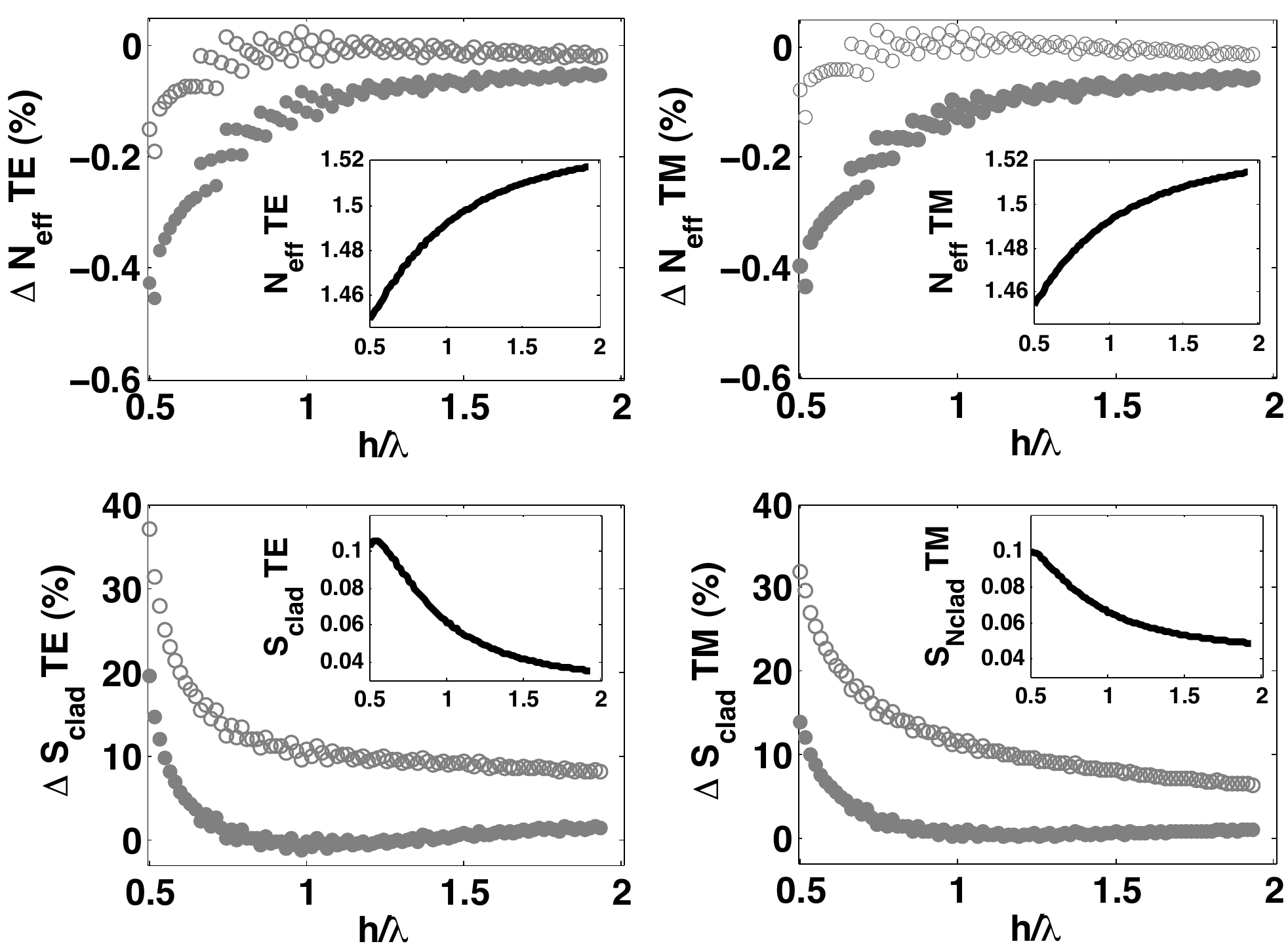}
\caption{Comparison of effective indices and sensitivites to the cladding refractive index for fundamental TE and TM modes of a typical polymeric rectangular waveguide with varying height. Reference ADI values, relative differences between ADI and EIM values, and between ADI and MM values are displayed in both instances. Legend: MM = disks, EIM = circles and ADI = solid (nested graphs).}
\label{fig:Figure5P1V1}
\end{center}\end{figure}

In both comparisons, we consider the ADI values to be the reference values. Results are gathered in the form of graphs in Fig.\ref{fig:Figure5P1V1}.

\subsection{Discussion}

The first noticeable and expected result of the comparison is that for both EIM and MM, the accuracy rapidly decreases with the height of the waveguide. In this case, the common assumption of the separation of spatial variables for the electromagnetic field becomes questionable since for a weakly-guided mode, a large part of the intensity is located in the corners of the waveguide where singular dielectric corner effects occur \cite{Hadley02}. 

It also seems that the additional assumption made in MM (cf. Annex C) is detrimental to the accuracy of the calculated effective index. This is well explained since the assumption that the electromagnetic field can be neglected in the corner regions is not valid for weakly-guided modes.

The effective indices calculated from both methods are close to the ones obtained with ADI, especially in the case of EIM with relative errors under 0.2$\%$. This shows that numerical methods can be effectively replaced in most cases by EIM when it comes to calculating effective index values. In addition, ADI method can be problematic to use in cases of weak guidance since electromagnetic field components are arbitrarily set to zero at the border of the simulation window, biasing the characteristics of the computed guided mode. This is particularly visible for the calculations of $S_{clad}$ discussed below. The only solution to overcome this difficulty is to increase the size of the simulation window for weakly guided modes while maintaining a constant grid density. This drastically increases computer resource consumption. We also point out that the visible discontinuous behaviors of both $\Delta N_{eff}TE$ and $\Delta N_{eff}TM$ near $h/\lambda=1$ are due to bias introduced by the ADI method, related to the limited density of the simulation grid.

When comparing sensitivities to the cladding refractive index. Data shows that MM provides sufficient accuracy for initial waveguide design, within 5\% of ADI estimated values for $h/\lambda$ ratios over 0.6, and within 2\% over 0.7.

It is particularly interesting to notice that for this sensitivity coefficient, MM provides more accurate results than EIM, for both TE and TM modes. Since calculated effective indexes are slightly more accurate with EIM, this result appears odd. However, it may be explained by the fact that as a derivative, the sensitivity potentially cancels systematic errors related to input effective indexes values (cf. Part 2).

The major discrepancy between analytical and numerical methods for $S_{clad}$ is near the cut-off (not represented), where the numerical method predicts a saturation of the cladding sensitivity. This effect comes from the finite aspect of the calculation window in the ADI method, and affects the accuracy of the cladding sensitivities more than the accuracy of the effective indexes. This adds to the limitations of the numerical methods for sensitivity studies.

\section{Application to the optimization of evanescent-wave sensors based on rectangular waveguides}
\subsection{Principle}
The reference performance for waveguide-based evanescent field sensors is the detection limit \cite{White08}, often defined as the minimal refractive index variation of the cladding that is detectable, i.e. that generates a signal higher than the noise level. With such definition, the detection limit is expressed as $DL = R/S_{clad}$
where $R$ is the smallest measurable variation of effective index, and $S_{clad}$ is the sentivity of the effective index to the cladding index. Although $R$ may depend on the waveguide itself, it is mainly limited by the lightsource, the photodetectors and other such devices. As a consequence, the principle of our optimization is to maximize $S_{clad}$.

Using our approach, the sensitivity $S_{clad}$ of a rectangular waveguide can be evaluated with good accuracy, and for a large number of configurations {\bf V} = ($N_{core}$, $N_{clad}$, $N_{sub}$, $w$, $h$,  $\lambda$) (cf. Part 4). Studies, similar to that published by Parriaux {\it et al.} regarding the optimization of the sensitivity of slab waveguides \cite{Parriaux98}, can therefore be undertaken for rectangular waveguides. In the following, we only focus on the influence of a few parameters on $S_{clad}$.

\subsection{Methodology}
\begin{figure}[here]
\includegraphics[width=7cm]{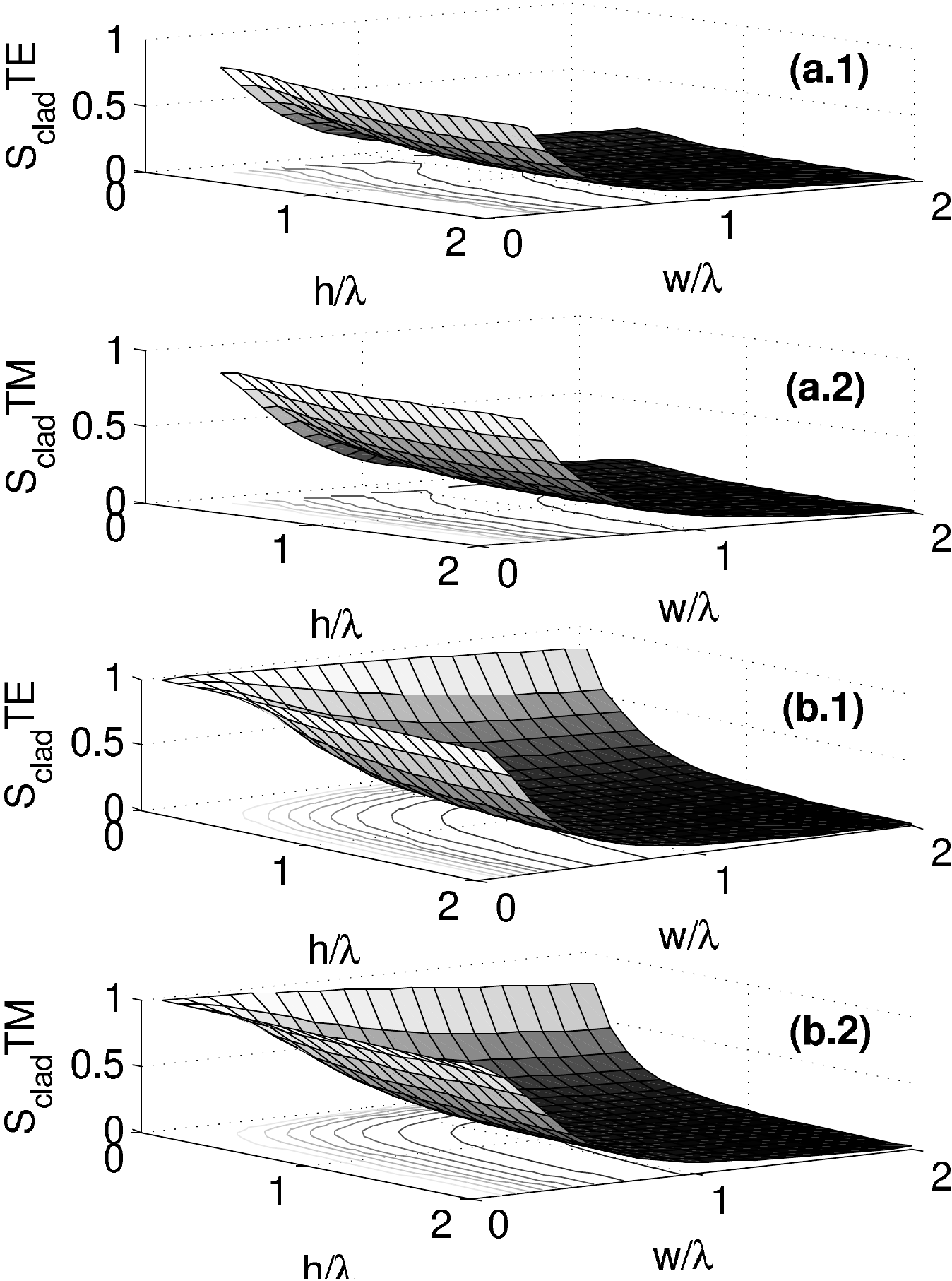}
\caption{Sensitivities of the effective indices of fundamental TE and TM modes to the refractive index of the cladding of a rectangular waveguide vs normalized height and width. (a) asymmetric waveguide with $N_{core}=1.56$, $N_{sub}=1.444$ and $N_{clad}=1.323$, (b) symmetric waveguide with $N_{core}=1.56$ and $N_{clad}=N_{sub}=1.333$}
\label{fig:FigureB}
\end{figure}

We evaluate $S_{clad}$ for the fundamental TE and TM modes of two rectangular waveguides, with height $h$ and width $w$ as optimization parameters. Since it is more accurate, MM is chosen as analytical method to describe the rectangular waveguide (cf. Annex C). The first rectangular waveguide is similar to that defined in Part.4. The second waveguide is characterized by a SU-8 core, and by an identical material for substrate and cladding with a refractive index of 1.333 at 1550 nm. The results, displayed in Fig.\ref{fig:FigureB}, show two interesting trends.

\subsection{Optimal substrate material}
For a given width and height, the second waveguide has larger cladding sensitivities than the first waveguide. We interpret this by the fact that, because the second waveguide has a smaller substrate index, a higher proportion of the modal energy propagates in the cladding material, hence the higher sensitivity to the cladding refractive index. Following this trend, an optimal value of $S_{clad}$ is theoretically obtained for a given waveguide width and height when $N_{sub}=1$. More practically, the substrate material should be chosen with a refractive index as low as possible in order to optimize $S_{clad}$.

\subsection{Optimal waveguide geometry}
For a given set of waveguide materials, it is clear from Fig.\ref{fig:FigureB} that both waveguide width $w$ and height $h$ are extremely important in the optimization of $S_{clad}$. As expected, maximal sensitivities are obtained for smaller values of $w$ and $h$. These maximal values correspond to the case where most of the modal energy actually propagates in the cladding and substrate regions rather than in the core waveguide.

\subsection{Synthesis}
A short trend analysis of the results displayed in Fig.\ref{fig:FigureB} show that in order to optimize the sensitivity $S_{clad}$ of a rectangular waveguide, and hence the detection limit $DL$ of an evanescent-wave sensor based on such waveguide, the substrate material should have a low refractive index, and the waveguide core should have a small cross-section. A finer analysis would have to take into account the various constraints due to available materials, waveguide fabrication and device operation. For such cases, our approach can still be efficiently used in more complex optimization schemes than the one we propose here.

\section{Optimization of the thermal sensitivity of integrated optical devices based on rectangular waveguides}
	\subsection{Principle}
External perturbations such as variations of pressure, humidity and temperature typically cause drifts in Integrated Optical (IO) devices operation.
For devices based on waveguides, this is often due to the fact that guided modes and their effective indexes are sensitive to these physical quantities. This is particularly true for temperature. 

As a consequence, most design efforts are usually aimed at reducing temperature variations of the device itself, for instance by thermal regulation of the device \cite{Delezoide12}. However, a more powerful method involves the reduction of device sensitivity to its own temperature. In this optmization scheme, the sensitivity $S_T$ of the effective indexes of the guided modes to the temperature $T$ should be minimal.

\subsection{Methodology}
 
In order to evaluate $S_T$, an approximate expression is commonly used \cite{Raghunathan10}:
\begin{equation}\label{5Ca}
S_T = \Gamma_{core}\frac{dN_{core}}{dT}+\Gamma_{clad}\frac{dN_{clad}}{dT}+\Gamma_{sub}\frac{dN_{sub}}{dT}
\end{equation}
where $\Gamma$ coefficients are fractional parts of the total modal intensity in the different regions of the waveguide.

The main issue with this formulation is the major assumption that temperature variations only modify effective indexes through thermo-optic effects.
Consequently, Eq.\eqref{5Ca} neglects effective index variations due to the thermal expansion of the waveguide core. Yet, large thermo-optic (TO) coefficients are intrinsically correlated to large thermal expansion (TEx) coefficients \cite{Tan98}. Another issue of Eq.~\eqref{5Ca}, is the necessary and non-trivial calculation of $\Gamma$ coefficients. Even for relatively simple structures such as rectangular waveguides, this requires numerical methods to determine the full electromagnetic field distribution, for the entire waveguide.

Because of the limitations of Eq.\eqref{5Ca}, we propose for the optimization of the thermal sensitivity a new way to express $S_T$ taking into account both TO and TEx contributions to the effective index variations, and where coefficients are easy to calculate using our approach. First, if we apply Eq.\eqref{3m} with $X=T$ to a rectangular waveguide, we obtain a locally-exact expression for $S_T$:
 \begin{equation}\label{3m2}
\begin{split}
S_T &= \frac{dN_{eff}}{dT}= S_{core}\cdot\frac{dN_{core}}{dT}+S_{sub}\cdot\frac{dN_{sub}}{dT}\\
   & \quad+S_{clad}\cdot\frac{dN_{clad}}{dT}+S_{w}\cdot\frac{dw}{dT}+S_{h}\cdot\frac{dh}{dT}
\end{split}
\end{equation}
In order to evaluate $S_T$ for various configurations {\bf V} = ($N_{core}$, $N_{clad}$, $N_{sub}$, $w$, $h$,  $\lambda$), values of $S_i$ coefficients are calculated using the same method applied in Parts 4 and 5. The issue is then to retrieve adequate values for TO and TEx related $(d\cdot/dT)$ coefficients. 

In the following, we consider a rectangular waveguide similar to that described in Part 4. For this waveguide, both core and substrate materials, SU-8 and silica, are solids whereas the cladding material, DI water, is in liquid phase. As a first approximation, we can assume that the core material is free to expand in all directions. This implies that linear TEx coefficients are similar for both waveguide width $w$ and height $h$. Therefore, we verify: 
\begin{equation}
\frac{1}{h}\frac{dh}{dT}=\frac{1}{w}\frac{dw}{dT} = \beta_{core}
\end{equation}
where $\beta_{core}$ is the linear TEx coefficient of the core material. With this simplification, we can express the thermal sensitivity $S_T$ as:
\begin{equation}\label{5Cb}\begin{split}
S_T &= S_{core}\cdot\alpha_{core}+S_{clad}\cdot\alpha_{clad}+S_{sub}\cdot\alpha_{sub}\\
&\qquad + \left( w\cdot S_{w} + h\cdot S_{h} \right)\beta_{core}
\end{split}\end{equation}
where $\alpha=(dN/dT)$ coefficients are thermo-optic coefficients. From \cite{Feng03}, \cite{Malitson65}, and \cite{Kamikawachi08} we gather the following values of these coefficients: $\alpha_{core}=-1.87\times10^{-4} RIU.K^{-1}$, $\alpha_{sub}=+1.28 \times 10^{-5} RIU.K^{-1}$, and $\alpha_{clad}=-8.0 \times 10^{-5} RIU.K^{-1}$. We also use $\beta_{core} = 152\; ppm.K^{-1}$ from \cite{Feng03}.

\subsection{Results and discussion}

\begin{figure}[h]
\begin{center}
\centerline{\includegraphics[width=0.9\columnwidth]{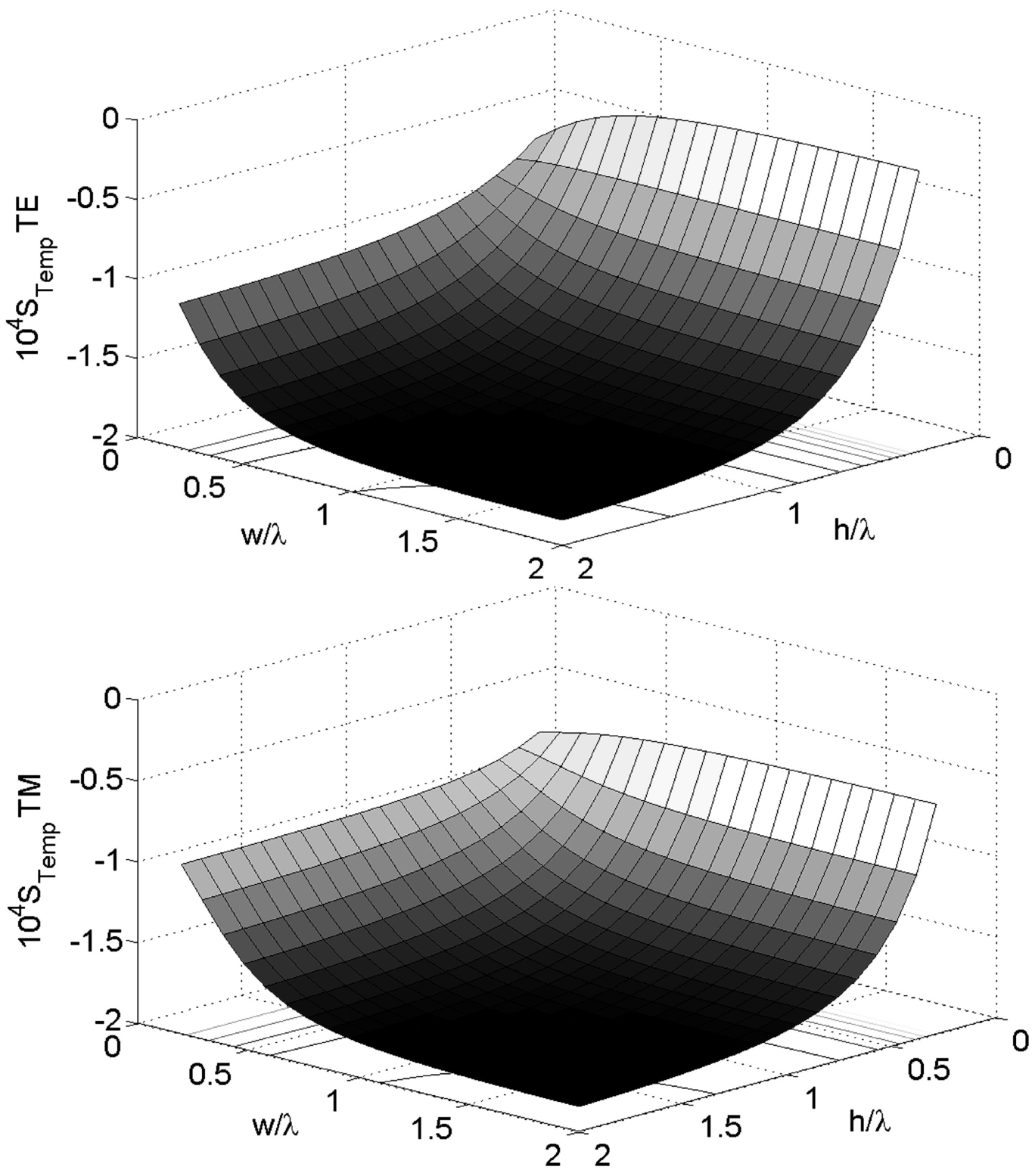}}
\caption{Full thermal sensitivities $S_{T}$ of the fundamental TE and TM modes to the temperature of a rectangular waveguide versus the normalized height and width.}
\label{fig:FigureD}
\end{center}
\end{figure}

Evaluations of the full thermal sensitivities of the fundamental TE and TM modes of the previously defined rectangular waveguide for varying height and width are presented in Fig.~\ref{fig:FigureD}.

The overall negative values of both thermal sensitivities show the dominant role of negative TO contributions due to the SU-8 core and aqueous cladding. More specifically, it is clear in Fig.~\ref{fig:FigureD} that the highest thermal sensitivities are obtained for large widths and heights, corresponding to the situations where the guided mode is well confined in the SU-8 core.

Because of the large TO coefficient of the core material, we show that whatever the height and width, the thermal sensitivity cannot be significantly reduced. However, if a core material with a lower TO coefficient was to be used, our approach would be useful to find adequate sets of waveguide parameters to minimize the thermal sensitivity.


Regarding the accuracy of such studies, we point out that, in order to obtain Eq.\eqref{5Cb}, it is assumed that the core waveguide is free to expand in both vertical and lateral directions. However, it is well known \cite{Zhao03} that, due to the mismatch of TEx coefficients between the core and the substrate, variations in temperature cause stress-induced birefringence. This phenomenon could be taken into account by applying a small correction to the $\beta$ coefficient in front of $S_{w}$ \cite{Feng03}.

The accuracy obtained in the determination of $S_T$ values from Eq.\eqref{5Cb} also depends on the accuracies of TEx and TO coefficients values available. Experimental values for most commonly used materials can generally be retrieved in the literature, but a full reliability of these values is not guaranteed due to the inherent variability of fabrication processes. 

\section{Evaluation of computing time and memory usage}

For a general optimization procedure of a rectangular waveguide, the optimization parameters are $N_{core}$, $N_{clad}$, $N_{sub}$, $w$, $h$, and $\lambda$. These parameters can be combined in a vector {\bf V} = ($N_{core}$, $N_{clad}$, $N_{sub}$, $w$, $h$,  $\lambda$) that characterizes one trial configuration. For one central configuration {\bf V$_0$}, the number of sensitivity coefficients $S_i$ to evaluate depends on the methodology of the optimization. 

Let us consider the case of thermal sensitivity optimization for a rectangular waveguide, as described in Part 6. Using Eq.\eqref{5Cb}, we need to calculate five $S_i$ coefficients for each thermal sensitivity evaluation.

Using a numerical method, each value of a single coefficient $S_i$ - the sensitivity of the effective index to the parameter $a_i$- requires the evaluation of at least two effective index values: $N_{eff}(\dots, a_i+\delta, \dots)$ and $N_{eff}(\dots, a_i-\delta, \dots)$ where $\delta$ is an increment (cf. Part 4).
Assuming that each independent parameter is sampled on $P$ points, and writing down $n$ as the number of independent optimization parameters, the number of independent configurations is $P^{n}$. Thus, in order to calculate the values of a single coefficient $S_i$ for all independent configurations, the total number of effective index evaluations to perform is $2\times P^{n}$.

Consequently, in our previous example, $10\times P^{5}$ effective index calculations are necessary to evaluate the thermal sensitivity of a rectangular waveguide - for which $n=5$ - for all independent configurations. It amounts to $31,250$ calculations for $P=5$, $10^6$ for $P=10$ and $3.125\times10^9$ for $P=50$. This has to be combined to the fact that, with our commercial mode solver using the ADI method, a precise calculation of a single effective index can take up to several minutes with a standard PC. Also, the high grid density required when small increments of the waveguide width or height are applied translate into a massive memory usage.

In comparison, when the methodology described in Part 6 is used, the same study only requires $P^{n}$ effective index calculations for each guided mode; ten times less than with a numerical method. This is due to the fact that sensitivity coefficients necessary to calculate the thermal sensitivity of a configuration {\bf V$_0$} are directly derived from a single effective index evaluation. Furthermore, using MM or EIM, the computation time for such an evaluation is down to a few milliseconds with a standard PC. The effective index evaluation being the most time-consuming step, the overall computation time is therefore greatly reduced when our approach is used. In addition, the memory usage when this procedure is applied is low.

\section{Generalization of the approach to all waveguides}

Let us consider a waveguide that can be physically described by a set of N parameters $\left( a_{1}, a_{2}, ... , a_{N} \right)$, and a guided mode of this waveguide described by an effective index $N_{eff}$. In analogy to the cases of the slab and rectangular waveguides, we can assume that the total derivative $dN_{eff}$ of this effective index can be written, using sensitivity coefficients $S_i$, as:
\begin{equation}
\label{2a}
dN_{eff} = \sum\limits_{i=1}^N \frac{\partial N_{eff}}{\partial a_{i}}da_{i}=\sum\limits_{i=1}^N S_{i}da_{i}
\end{equation}

We have shown in Part 1 and Part 6 that, provided the expressions of all $S_i$ coefficients are known, the sensitivity $S_X$ of the effective index to a quantity $X$ is easily evaluated. Therefore, the issue of generalizing the approach comes down to finding a general method to express all $S_i$ coefficients for a given waveguide.

In the various cases we studied throughout this article, we manage to evaluate the effective index $N_{eff}$ after the resolution of a finite number $Q$ of equations. In addition, we noticed that when we apply a total derivative to these $Q$ equations, we are able to identify an expression for each sensitivity coefficient after simple arithmetical operations.

Depending on the waveguide, and the analytical method used to describe it, the number $M$ of equations to consider varies. For instance, we show that $Q=1$ for the slab waveguide (cf. Part 1), $Q=2$ for the rectangular waveguide using the EIM (cf. Annex B), and $Q=3$ for the rectangular waveguide using MM (cf. Annex C).

It seems reasonable to assume that, with adequate analytical models, the same rules apply to most waveguides: rib or ridge waveguides, step index optical fibers, plasmonic waveguides, and even more complex waveguides. Therefore, our approach can potentially apply to study the effective index sensitivities of a wide variety of waveguides.

However, we have shown in Part.4 that the accuracy of the results is highly correlated to the analytical method chosen to describe the waveguide. This suggests that our approach is not suited to study the sensitivities of waveguides which are difficult to model analytically.
\section{Conclusion}

In this article, we have introduced a new, efficient approach for waveguide sensitivity analysis and optimization.
We tested this approach by applying it to the case of the rectangular waveguide, and found sensitivity values in good agreement - when Marcatili's method (MM) is employed- with that obtained using a fully-numerical approach. Despite the lesser accuracy of our approach, it allows a significant improvement over computing time and memory usage. In our opinion, this makes our approach particularly suited for waveguide sensitivity analysis and optimization, at least in early design stages of the waveguide. It can thereby be useful for many applications, among which waveguide design for the optimization of evanescent-wave sensors.

Moreover, the mathematical formalism developed within our approach makes it relatively easy to satisfactorily express and evaluate quantities such as the waveguide sensitivity to temperature, as described in Part 6, but also the waveguide sensitivity to mechanical constraints, and to other physical quantities that affect simultaneously most waveguide parameters.

Finally, even though we only applied our approach to the slab and rectangular waveguides in this article, we discussed the possibility of applying it to perform sensitivity analysis and optimization for virtually any waveguide, as long as it can be satisfactorily described by a finite number of analytical dispersion equations.

\appendix

\section{Dispersion equations of the slab waveguide}

In order to write the dispersion equations of the slab waveguide in a form that is easily solvable, we introduce three normalized dimensionless parameters: the normalized propagation constant $b$, the normalized frequency $v$ and the asymmetry factor $\gamma$, defined herein as:
\begin{equation}\label{3}
\left\{
\begin{aligned}
&b=\frac{N_{eff}^2-N_{sub}^2}{N_{core}^2-N_{sub}^2} \; ; \; \gamma=\frac{N_{sub}^2-N_{clad}^2}{N_{core}^2-N_{sub}^2} \\
&v=kw \left ( N_{core}^2-N_{sub}^2 \right )^{1/2}
\end{aligned}
\right.
\end{equation}
where k is the wavenumber in vacuum.

With $b$, $v$ and $\gamma$ thus defined, using a method similar to that described in \cite{Okamoto00}, we obtain two families of identically zero functions, $f_{TE}(m)$ and $f_{TM}(m)$, expressed as: 
\begin{equation}
\left\{
\begin{aligned}
\label{3d}
f_{TE}(m) &= 0=v \left(1-b\right)^{1/2} - m\pi \\
& - \tan^{-1} \left( \frac{b}{1-b} \right)^{1/2} - \tan^{-1} \left( \frac{b+\gamma}{1-b} \right)^{1/2}\\
f_{TM}(m) &= 0= v \left(1-b\right)^{1/2} - m\pi \\
&\quad- \tan^{-1} \left( \frac{N_{core}^{2}}{N_{sub}^{2}} \left( \frac{b}{1-b} \right)^{1/2} \right) \\
&\quad \quad - \tan^{-1} \left( \frac{N_{core}^{2}}{N_{clad}^{2}} \left( \frac{b+\gamma}{1-b} \right)^{1/2} \right)
\end{aligned} 
\right.
\end{equation}
In the following, above equations are used as dispersion equations for all TE and TM guided modes, respectively. In those equations,  we define the positive integer $m$ as the order of the corresponding transverse mode, m=0 corresponding to the fundamental mode for both polarizations.

We assumed here that the substrate index is larger than the cladding index. In the opposite situation where $N_{clad} > N_{sub}$, both variables must be switched in the above definitions and in the dispersion equations.

\section{Sensitivity coefficients for a rectangular waveguide using the Effective Index Method}
\subsection{Description of the Effective Index Method}
In the Effective Index Method (EIM), both transverse directions of optical confinement are treated sequentially. The method is based on the assumption that the electromagnetic field can be expressed with a separation of the spatial variables \cite{Okamoto00}. In the case of a rectangular waveguide, a confinement direction - lateral or vertical - is arbitrarily chosen. Along the chosen direction, a first dispersion equation provides the partial effective index $N_{eff,\;I}$. In the opposite direction, the rectangular waveguide is treated as a slab waveguide with $N_{eff,\;I}$ as core refractive index. A second dispersion equation finally provides the effective index of the studied guided mode.

\subsection{Dispersion equations}
In the following, we choose to start with the vertical direction. In analogy to the method applied to solve the slab waveguide, we need to introduce the normalized propagation constant $b_{V}$ and the normalized frequency $v_{V}$, both related to the vertical direction. We also need to introduce their equivalent $b_{L}$ and $v_{L}$ for the lateral direction.
\begin{equation}\label{4A2a}
\left\{
\begin{aligned}
b_{V}&=\frac{N_{eff, \;I}^2-N_{sub}^2}{N_{core}^2-N_{sub}^2} \; ; \; v_{V}=kh \left ( N_{core}^2-N_{sub}^2 \right )^{1/2}\\
b_{L}&=\frac{N_{eff}^2-N_{clad}^2}{N_{eff, \;I}^2-N_{clad}^2}\; ; \; v_{L}=kw \left (N_{eff, \;I}^2-N_{clad}^2 \right )^{1/2}
\end{aligned}
\right.
\end{equation}

Applying the method described in \cite{Okamoto00}, we find that guided modes of rectangular waveguides are determined by polarization and mode-order dependent sets of two identically zeros functions, expressed below for both quasi-TM and quasi-TE modes.

\subsubsection{Dispersion equations for Quasi-TM $\left(E_{p,q}^{y}\right)$ modes}

\begin{equation}\label{4A2e}
\left\{
\begin{aligned}
f_{L}^{(p)} &= 0=v_{L}\left(1-b_{L}\right)^{1/2} - p\pi - 2\tan^{-1} \left( \frac{b_{L}}{1-b_{L}} \right)^{1/2}\\
f_{V}^{(q)} &= 0 =v_{V} \left(1-b_{V}\right)^{1/2} - q\pi \\
&\qquad- \tan^{-1} \left( \frac{N_{core}^{2}}{N_{sub}^{2}} \left( \frac{b_{V}}{1-b_{V}} \right)^{1/2} \right) \\
&\qquad \qquad - \tan^{-1} \left( \frac{N_{core}^{2}}{N_{clad}^{2}} \left( \frac{b_{V}+\gamma}{1-b_{V}} \right)^{1/2} \right)
\end{aligned}
\right.
\end{equation}

\subsubsection{Dispersion equations for Quasi-TE $\left(E_{p,q}^{x}\right)$ modes}

\begin{equation}\label{4A2g}
\left\{
\begin{aligned}
f_{L}^{(p)} &=0= v_{L}\left(1-b_{L}\right)^{1/2} - p\pi \\
&\qquad \qquad- 2\tan^{-1} \frac{N_{eff, \;I}^2} {N_{clad}^{2}}\left(\frac{b_{L}}{1-b_{L}} \right)^{1/2}\\ 
f_{V}^{(q)} &=0= v_{V} \left(1-b_{V}\right)^{1/2} - q\pi - \tan^{-1} \left(\frac{b_{V}}{1-b_{V}} \right)^{1/2}\\
&\qquad \qquad \qquad -\tan^{-1} \left(\frac{b_{V}+\gamma}{1-b_{V}} \right)^{1/2}
\end{aligned}
\right.
\end{equation}

In these dispersion equations, we define the positive integers $p$ and $q$ as the lateral and vertical mode orders, respectively. We also define $E^x_{0,0}$ and $E^y_{0,0}$ as the fundamental quasi-TM and quasi-TE modes, respectively. This means that the dominant electromagnetic component of $E^x_{p,q}$ and $E^y_{p,q}$ modes have $p$ zeros in the lateral direction and $q$ zeros in the vertical direction.\\

We point out that we assumed $N_{clad}<N_{sub}$. Due to the fact that the lateral direction of confinement always relate to a symmetric slab waveguide, the case where the cladding index is larger than the substrate index can be treated simply by switching $N_{clad}$ and $N_{sub}$ in all expressions related to the vertical - asymmetrical - direction: $b_{V}$, $v_{V}$, $f_{V}$, $\gamma$, $f_{L}^{(p)}$ and $f_{V}^{(q)}$.

\subsection{Formulas for the $S_i$ coefficients}
Similarly to the case of the slab waveguide, the derivation of a set of dispersion equations, valid for a given mode, determines the relation between the quantity $dN_{eff}$ and the elementary variations of all other parameters of the rectangular waveguide:
\begin{equation}\label{4A3a}
\begin{split}
dN_{eff}  &= S_{core}dN_{core}+S_{sub}dN_{sub}+S_{clad}dN_{clad} \\
& \quad \quad+S_{w}dw+S_{h}dh+S_{\lambda}d\lambda
\end{split}
\end{equation}

The difference with the case of the slab waveguide is that, in order to obtain the expressions of the $S_i$ coefficients, the differentiation of two dispersion equations, $f_{V}$ and $f_{L}$, is necessary. It is the reason why the expressions of the $S_i$ coefficients are somewhat different in form, with one or two components depending on the coefficient.

From the differentiation of dispersion equations sets \eqref{4A2e} or \eqref{4A2g}, we obtain the formulas for $S$ coefficients related to the waveguide refractive indices:
\begin{equation}\label{4A3b}
\left\{
\begin{aligned}
&S_{core} = \left(\frac{\partial f_{L}}{\partial N_{eff}} \right)^{-1}\cdot\frac{\partial f_{L}}{\partial N_{eff, \;I}} \\
& \qquad \qquad \qquad \qquad \times \left(\frac{\partial f_{V}}{\partial N_{eff,\;I}} \right)^{-1}\cdot\frac{\partial f_{V}}{\partial N_{core}} \\
&S_{sub} = \left(\frac{\partial f_{L}}{\partial N_{eff}} \right)^{-1}\cdot\frac{\partial f_{L}}{\partial N_{eff, \;I}} \\
& \qquad \qquad \qquad \qquad \times\left(\frac{\partial f_{V}}{\partial N_{eff,\;I}} \right)^{-1}\cdot\frac{\partial f_{V}}{\partial N_{sub}}  \\
&S_{clad} =\left(\frac{\partial f_{L}}{\partial N_{eff}} \right)^{-1}\cdot\frac{\partial f_{L}}{\partial N_{eff, \;I}} \\
& \qquad \qquad \qquad \qquad \times \left(\frac{\partial f_{V}}{\partial N_{eff,\;I}} \right)^{-1}\cdot\frac{\partial f_{V}}{\partial N_{clad}}\\
&\qquad-\left(\frac{\partial f_{L}}{\partial N_{eff}} \right)^{-1}\cdot\frac{\partial f_{L}}{\partial N_{clad}}
\end{aligned}
\right.
\end{equation}
The expression of $S_{clad}$ contrasts with those of $S_{sub}$ and $S_{core}$ because unlike $N_{sub}$ and $N_{core}$, $N_{clad}$ is an explicit variable of both functions $f_V$ and $f_L$, hence the additional term.

From the same procedure, we also obtain the formulas for $S$ coefficients related to the geometry of the rectangular waveguide:
\begin{equation}\label{4A3bbis}
\left\{
\begin{aligned}
&S_{h} = \left(\frac{\partial f_{L}}{\partial N_{eff}} \right)^{-1}\cdot\frac{\partial f_{L}}{\partial N_{eff, \;I}}\\
&\qquad \qquad \qquad \qquad  \times \left(\frac{\partial f_{V}}{\partial N_{eff,\;I}} \right)^{-1}\cdot\frac{\partial f_{V}}{\partial h}\\
&S_{w} = -\left(\frac{\partial f_{L}}{\partial N_{eff}} \right)^{-1} \cdot \frac{\partial f_{L}}{\partial w} 
\end{aligned}
\right.
\end{equation}
It also possible to express the dependency of the effective index to the wavelength from the $S_\lambda$ coefficient:
\begin{equation}\label{4A3c}
\left\{
\begin{aligned}
&S_{\lambda} =\left(\frac{\partial f_{L}}{\partial N_{eff}} \right)^{-1}\frac{\partial f_{L}}{\partial N_{eff, \;I}} \\
& \times \left(\frac{\partial f_{V}}{\partial N_{eff,\;I}} \right)^{-1}\frac{\partial f_{V}}{\partial \lambda} -\left(\frac{\partial f_{L}}{\partial N_{eff}} \right)^{-1}\frac{\partial f_{L}}{\partial \lambda}
\end{aligned}
\right.
\end{equation}
Using these formulas, each $S_i$ coefficient can be expressed as a function of normalized parameters $\left( b_V, b_L, v_V, v_L, \gamma \right)$ or as a function of physical parameters $\left( N_{eff}, N_{eff,I}, N_{core}, N_{clad}, N_{sub}, w, h, \lambda \right)$.
\section{Sensitivity coefficients for a rectangular waveguide using Marcatili's method}
\subsection{Description}
In Marcatili's method (MM), both directions of optical confinement are treated independently. The important assumption of this method is that the electromagnetic field in the corner regions of the structure can be neglected, since it decays quite rapidly \cite{Marcatili69}. The separation of variables, used in EIM, is also assumed in MM. As a result, the effective index can be obtained by solving two independent 1D dispersion equation, one for the lateral confinement, providing $N_{eff,\;L}$, and one for the vertical confinement, providing $N_{eff,\;V}$.
The effective index of the guided mode is then determined by:
\begin{equation}
\label{4B1a}
N_{eff}^{2}= N_{eff,\;L}^{2}+N_{eff,\;V}^{2}- N_{core}^{2} 
\end{equation}
\subsection{Dispersion equations}
In the case of the rectangular waveguide we want to solve, we will need, in analogy to the slab waveguide treatment, the following dimensionless parameters:
\begin{equation}\label{4B2a}
\left\{
\begin{aligned}
b_{L}&=\frac{N_{eff,\;L}^2-N_{clad}^2}{N_{core}^2-N_{clad}^2} \; ; \; v_{L}=kw \left ( N_{core}^2-N_{clad}^2 \right )^{1/2}\\
b_{V}&=\frac{N_{eff,\;V}^2-N_{sub}^2}{N_{core}^2-N_{sub}^2}  \; ; \;v_{V}=kh \left ( N_{core}^2-N_{sub}^2 \right )^{1/2}
\end{aligned}
\right.
\end{equation}
The asymmetry factor $\gamma$ is defined the same way as in Eq.\eqref{3}.
With this set of parameters, the guided modes of the rectangular waveguides are solutions of a set of two equations which depends on the mode polarization.

\subsubsection{Dispersion equation for Quasi-TM $\left(E_{p,q}^{y}\right)$ mode}

\begin{equation}\label{4B2e}
\left\{
\begin{aligned}
&f_{L}(p) = 0 = v_{L}\left(1-b_{L}\right)^{1/2} - p\pi - 2\tan^{-1} \left( \frac{b_{L}}{1-b_{L}} \right)^{1/2}\\
&f_{V}(q) = 0 = v_{V} \left(1-b_{V}\right)^{1/2} - q\pi \\
&\qquad \qquad- \tan^{-1} \left( \frac{N_{core}^{2}}{N_{sub}^{2}} \left( \frac{b_{V}}{1-b_{V}} \right)^{1/2} \right) \\
&\qquad \qquad \qquad- \tan^{-1} \left( \frac{N_{core}^{2}}{N_{clad}^{2}} \left( \frac{b_{V}+\gamma}{1-b_{V}} \right)^{1/2} \right)
\end{aligned}
\right.
\end{equation}

\subsubsection{Dispersion equation for Quasi-TE $\left(E_{p,q}^{x}\right)$ mode}
\begin{equation}\label{4B2g}
\left\{
\begin{aligned}
&f_{L}(p) = v_{L}\left(1-b_{L}\right)^{1/2} - p\pi- 2\tan^{-1} \frac{N_{core}^{2}} {N_{clad}^{2}}\left(\frac{b_{L}}{1-b_{L}} \right)^{1/2}\\
&f_{V}(q) = v_{V} \left(1-b_{V}\right)^{1/2} - q\pi - \tan^{-1} \left(\frac{b_{V}}{1-b_{V}} \right)^{1/2} \\
&\qquad \qquad \qquad- \tan^{-1} \left(\frac{b_{V}+\gamma}{1-b_{V}} \right)^{1/2}
\end{aligned}
\right.
\end{equation}

\subsection{Formulas for the $S_i$ coefficients}
The total derivative $dN_{eff}$ is calculated similarly to the case of the effective index method and provides the formulas for the S coefficients, first related to the refractive indices:

\begin{equation}\label{4B3a}
\left\{
\begin{aligned}
&S_{N_{core}} =-\frac{N_{eff,\;L}}{N_{eff}}\left(\frac{\partial f_{L}}{\partial N_{eff,\;L}} \right)^{-1}\frac{\partial f_{L}}{\partial N_{core}} \\
&\quad -\frac{N_{eff,\;V}}{N_{eff}}\left(\frac{\partial f_{V}}{\partial N_{eff,\;V}} \right)^{-1}\frac{\partial f_{V}}{\partial N_{core}} -\frac{N_{core}}{N_{eff}} \\
&S_{N_{clad}} =-\frac{N_{eff,\;L}}{N_{eff}}\left(\frac{\partial f_{L}}{\partial N_{eff,\;L}} \right)^{-1}\frac{\partial f_{L}}{\partial N_{clad}} \\
&\qquad\qquad\quad-\frac{N_{eff,\;V}}{N_{eff}}\left(\frac{\partial f_{V}}{\partial N_{eff,\;V}} \right)^{-1}\frac{\partial f_{V}}{\partial N_{clad}} \\
&S_{N_{sub}} =-\frac{N_{eff,\;V}}{N_{eff}}\left(\frac{\partial f_{V}}{\partial N_{eff,\;V}} \right)^{-1}\frac{\partial f_{V}}{\partial N_{sub}} 
\end{aligned}
\right.
\end{equation}
Then, those related to the geometry can be written as:
\begin{equation}\label{4B3abis}
\left\{
\begin{aligned}
&S_{w} =-\frac{N_{eff,\;L}}{N_{eff}}\left(\frac{\partial f_{L}}{\partial N_{eff,\;L}} \right)^{-1}\frac{\partial f_{L}}{\partial w} \\
&S_{h} =-\frac{N_{eff,\;V}}{N_{eff}}\left(\frac{\partial f_{V}}{\partial N_{eff,\;V}} \right)^{-1}\frac{\partial f_{V}}{\partial h}
\end{aligned}
\right.
\end{equation}
Finally, the sensitivity coefficient related to wavelength-dependency is expressed as:
\begin{equation}\label{4B3ater}
\left\{
\begin{aligned}
&S_{\lambda} = -\frac{N_{eff,\;L}}{N_{eff}}\left(\frac{\partial f_{L}}{\partial N_{eff,\;L}} \right)^{-1}\frac{\partial f_{L}}{\partial \lambda} \\
&\qquad\qquad\qquad-\frac{N_{eff,\;V}}{N_{eff}}\left(\frac{\partial f_{V}}{\partial N_{eff,\;V}} \right)^{-1}\frac{\partial f_{V}}{\partial \lambda} 
\end{aligned}
\right.
\end{equation}

\subsection{Limitation of Marcatili's method}
It can be observed that near cutoff, calculated values of the effective index can become lower than the lowest refractive index of the structure. This absurd behavior is attributed to the strong assumption that the electromagnetic field can be neglected in the corner regions, whose validity is seriously challenged when most of the mode intensity is propagating outside the core. However, when the effective index remains larger than the lowest refractive index, accuracy of the method is sufficient for waveguide analysis.

\end{document}